\documentstyle[aasms4, 11pt]{article}
\singlespace
\begin{document}

\def\gsim { \lower .75ex \hbox{$\sim$} \llap{\raise .27ex \hbox{$>$}} }
\def\lsim { \lower .75ex \hbox{$\sim$} \llap{\raise .27ex \hbox{$<$}} }
\title{The core density of dark matter halos: a critical challenge to the
$\Lambda$CDM paradigm?}
\lefthead{Navarro \& Steinmetz}
\righthead{Fatal problem to $\Lambda$CDM paradigm}

\author{Julio F. Navarro\altaffilmark{1}}

\affil{Department of Physics and Astronomy, University of
Victoria, Victoria, BC V8P 1A1, Canada}

\and 

\author{Matthias Steinmetz\altaffilmark{2}}

\affil{Steward Observatory, University of Arizona, Tucson, AZ 85721, USA}

\altaffiltext{1}{CIAR Scholar \& Alfred P.~Sloan Fellow. Email: jfn@uvic.ca}
\altaffiltext{2}{Alfred P.~Sloan Fellow \& David and Lucile Packard Fellow. 
Email: msteinmetz@as.arizona.edu}

\begin{abstract}
We compare the central mass concentration of Cold Dark Matter halos found in
cosmological N-body simulations with constraints derived from the Milky Way disk
dynamics and from the Tully-Fisher relation. For currently favored values of the
cosmological parameters ($\Omega_0 \sim 0.3$; $\Lambda_0=1-\Omega_0 \sim 0.7$;
$h \sim 0.7$; COBE- and cluster abundance-normalized $\sigma_8$; Big-Bang
nucleosynthesis $\Omega_b$), we find that halos with circular velocities
comparable to the rotation speed of the Galaxy have typically {\it three times}
more dark matter inside the solar circle than inferred from observations of
Galactic dynamics. Such high central concentrations of dark matter on the scale
of galaxy disks also imply that stellar mass-to-light ratios much lower than
expected from population synthesis models must be assumed in order to reproduce
the zero-point of the Tully-Fisher relation. Indeed, even under the extreme
assumption that {\it all} baryons in a dark halo are turned into stars, disks
with conventional $I$-band stellar mass-to-light ratios ($M/L_I \sim 2 \pm 1
(M/L_I)_{\odot}$) are about two magnitudes fainter than observed at a given
rotation speed. We examine several modifications to the $\Lambda$CDM model that
may account for these discrepancies and conclude that agreement can only be
accomplished at the expense of renouncing other major successes of the
model. Reproducing the observed properties of disk galaxies thus appears to
demand substantial revision to the currently most successful model of structure
formation.
\end{abstract}


\section{Introduction}

Over the past few years, cosmological models based on the paradigm of an
inflationary universe dominated by cold dark matter (CDM) have proved remarkably
successful at explaining the origin and evolution of structure in the
Universe. The free parameters of astrophysical relevance in this modeling are
surprisingly few: the current rate of universal expansion, $H_0$; the mass
density parameter, $\Omega_0$; the primordial baryon abundance, $\Omega_b$; and
the overall normalization of the power spectrum of initial density fluctuations,
$\sigma_8$. Over the past few years, limits on the values allowed for these
parameters have been consistently refined by improved observational techniques
and theoretical insight, and it is widely accepted that a new ``standard'' model
has emerged as the clear front-runner amongst competing models of structure
formation.

This model, which we shall call ``standard'' $\Lambda$CDM, or s$\Lambda$CDM for
short, envisions an eternally expanding universe with the following properties
(Bahcall et al 1999): (i) matter makes up at present less than about a third of
the critical density for closure ($\Omega_0 \sim 0.3$); (ii) a non-zero
cosmological constant restores the flat geometry predicted by most inflationary
models of the early universe ($\Lambda_0=1-\Omega_0\sim 0.7$); (iii) the present
rate of universal expansion is $H_0 \sim 70$ km s$^{-1}$ Mpc$^{-1}$ ($h=H_0/100$
km s$^{-1}$ Mpc$^{-1} \sim 0.7$); (iv) baryons make up a very small fraction of
the mass of the universe ($\Omega_b \approx 0.0125 \, h^{-2} \sim 0.0255 \ll
\Omega_0$); and (v) the present-day rms mass fluctuations on spheres of radius 8
$h^{-1}$ Mpc is of order unity ($\sigma_8 \sim 1.1$). The s$\Lambda$CDM model is
consistent with an impressive array of well-established fundamental
observations: the age of the universe as measured from the oldest stars (e.g.,
Chaboyer et al 1998), the extragalactic distance scale as measured by distant
Cepheids (e.g., Madore et al 1998); the primordial abundance of the light
elements (e.g., Schramm \& Turner 1998), the baryonic mass fraction of galaxy
clusters (e.g., White et al 1993), the amplitude of the Cosmic Microwave
Background fluctuations measured by COBE (e.g., Lawrence, Scott \& White 1999),
the present-day abundance of massive galaxy clusters (e.g., Eke, Cole \& Frenk
1996), the shape and amplitude of galaxy clustering patterns (e.g., Wu, Lahav \&
Rees 1999), the magnitude of large-scale coherent motions of galaxy systems
(e.g., Strauss \& Willick 1995, Zaroubi et al 1997), and the world geometry
inferred from observations of distant type Ia supernovae (e.g., Perlmutter et al
1999, Garnavich et al 1998), among others.

Because its major parameters are fixed, s$\Lambda$CDM is an eminently
falsifiable model whose predictive power may be used to ascertain its validity
on scales different from those used to tune the model. One scale of particular
interest is that of individual galaxies, since few observational constraints on
these small scales have been used to adjust the parameters of s$\Lambda$CDM.
This exercise is especially compelling because the dark halo structure found
in cosmological N-body simulations of Cold Dark Matter universes seems at odds with
dynamical studies of disk galaxies that assign a substantial gravitational role
to the disk component (see, e.g., the ``maximal disk'' solutions of Debattista
\& Sellwood 1998, and references therein) as well as with rotation curve studies
of dark matter-dominated galaxies (Moore 1994, Flores \& Primack 1994, McGaugh
\& De Block 1998, Moore et al 1999, Navarro 1999). These claims are based
largely on comparisons of the detailed {\it shape} of the rotation curve of very
low surface brightness dwarfs with the innermost {\it density profile} of
simulated dark halos. Unfortunately, the scales where deviations are most
pronounced (the inner few kpc) are also the most compromised by numerical
uncertainties (most simulations relevant to this problem published to date have
gravitational softening scales of order $1$-$2$ kpc). The comparison is thus
rather uncertain. For example, Kravtsov et al (1998) have argued, on the basis
of simulations similar to those used by the other authors, that CDM halos are
actually {\it consistent} with the rotation curves of dark matter-dominated
disks, a somewhat surprising result that illustrates well, nonetheless, the
vulnerability of numerical techniques on scales close to the numerical
resolution of the simulations. From a strictly pragmatic numerical standpoint,
it would be desirable to circumvent these uncertainties by adopting comparison
criteria that are less sensitive to the effects of numerical shortcomings.

One possible choice is to use, rather than the dark matter density profile near
the center, the {\it total amount} of dark mass within the main body of
individual galaxies. For spiral galaxies, this criterion would imply that
simulations that can estimate reliably the amount of dark mass within a couple
of exponential scalelengths may be safely used for comparison with
observations. For bright spirals like the Milky Way this corresponds to radii of
about $5$-$10$ kpc, well outside the region that may be compromised by numerical
artifacts in the current generation of N-body experiments.

In this paper we follow this proposal and compare the results of recent very
high-resolution simulations of the formation of dark halos in the s$\Lambda$CDM
model with observational constraints on the total dark mass within spiral disks
derived from the Tully-Fisher relation and from observations of Galactic
dynamics. The numerical setup of the simulations is identical to that described
by Navarro, Frenk \& White (1997, hereafter NFW97) but the number of particles
has been increased more than tenfold. As a result, each simulated s$\Lambda$CDM
halo has of order $250,000$ dark matter particles within the virial radius and
several thousands within radii comparable to the Sun's distance from the
Galactic center (the ``solar circle'' $R_{o}$) so our numerical uncertainties
are for all practical purposes negligible.

We show below that comparison between these simulations and observational
constraints reveals a severe inconsistency: s$\Lambda$CDM halos have
substantially more mass near the center than the maximum inferred from
observations. We argue that this presents a serious challenge to the
s$\Lambda$CDM cosmogony and to many of its likely variants and that possible
solutions involve fundamental revision of the basic premises or parameters of
the model.

\section{Observational Constraints on Dark Mass in Individual Galaxies}

\subsection{The Milky Way}
Kinematic observations of stars and gas in the Galaxy provide tight constraints
on the total amount of dark matter within the solar circle, $R_{o}$. A
direct estimate can be made by assuming that the halo is spherically symmetric,
$M_{\rm dark}(r<R_{o})= V_{\rm dark}^2(R_{o}) \, R_{o}/G$, where
$V_{\rm dark}(R_{o})$ is the contribution of the dark halo to the circular
velocity at $R_{o}$. This may be obtained by subtracting the disk
contribution from the total circular velocity, $V_{\rm
dark}^2(R_{o})=V_{c}^2(R_{o})-V_{\rm disk}^2(R_{o})$. For the
IAU-sanctioned values of $R_{o}=8.5$ kpc and $V_{\rm c}(R_{o})=220$ km
s$^{-1}$, and assuming that the disk potential is well approximated by an
exponential disk with scalelength $r_{\rm disk}=3.5$ kpc and total mass $M_{\rm
disk}= 6 \times 10^{10} M_{\odot}$ (Binney \& Tremaine 1987), we find
$$
M_{\rm dark}(r<R_{o})=5.2 \times 10^{10} M_{\odot}. \eqno(1)
$$
The uncertainty in this determination is hard to assess, although the evidence
suggests that the mass in eq.~1 is actually an {\it upper} limit to the dark
mass inside the solar circle. This is in good agreement with the recent Milky
Way mass models of Dehnen \& Binney (1998), who find that, within the slightly
larger radius of $10$ kpc the dark halo accounts for less than about $5$-$6
\times 10^{10} M_{\odot}$, and perhaps as little as $2.5 \times 10^{10}
M_{\odot}$. 

The disk contribution to the circular velocity increases with our estimated
value of $M_{\rm disk}$, which in turn depends on: (i) the local density of the
disk derived from the vertical kinematics of stars in the solar neighborhood
(i.e., from ``Oort limit'' analysis, $\Sigma_{\odot}\sim 70 M_{\odot} {\rm
pc}^{-2}$), (ii) on the exponential scalelength of the disk, and (iii) on the
solar circle itself, through $M_{\rm disk} \propto \Sigma_{\odot}\, r_{\rm
disk}^2 \,e^{R_{o}/r_{\rm disk}}$.  Recent reviews (Reid 1993, Sackett 1997) of
available data suggest that the exponential scalelength assumed above may need
to be revised downwards by up to $20\%$, but otherwise leave $\Sigma_{\odot}$
and $R_{o}$ largely unchanged\footnote{Note, however, that Olling \& Merrifield
(1998), however, argue for $R_{o}\sim 7.1$ kpc upon analysis of the role of
interstellar gas on the local values of Oort's constants. Since the circular
velocity at $R_{o}$ decreases correspondingly, this modification has little
effect on our conclusions.} from the values assumed above.  Such revision would
{\it increase} the disk mass, leading to values of $M_{\rm dark}(r<R_{o})$ lower
than quoted in eq.~1. Indeed, Sackett (1997) concludes that the Milky Way disk
may very well be ``maximal'' once this revision is taken into account. Finally,
we note that our procedure neglects the contribution of the Galaxy's bulge,
lending further support to our interpretation of eq.~1 as an upper limit to the
dark mass inside $R_{o}$.

Figure 1 compares, as a function of halo mass, the dark mass estimate in eq.~1
with the results of simulations of several s$\Lambda$CDM halos.  Halo masses
($M_{200}$) are measured inside the radius, $r_{200}$, of a sphere of mean
density 200 times the critical density for closure, and are typically
characterized by the circular velocity at that radius,
$V_{200}=(GM_{200}/r_{200})^{1/2} = (10\,G\,H_0\,M_{200})^{1/3}$. The reason for
this choice is that at $r_{200}$ the circular orbit timescale is approximately
equal to the age of the universe; $r_{200}$ thus separates the ``virialized''
region of the halo from the region where mass shells are infalling into the
system for the first time. The relevance of this definition stems from the
latter property: only baryons inside $r_{200}$ may contribute to the baryonic
mass of the galaxy, since those beyond this radius have yet to reach the center
of the halo (White et al 1993).

This property may be used to derive a firm {\it lower limit} to the mass of the
halo that surrounds the Milky Way, corresponding to the case where the baryonic
mass of the disk equals the total baryonic mass inside $r_{200}$: i.e., $M_{\rm
disk}\le M_{\rm disk}^{\rm max}=(\Omega_b/\Omega_0) M_{200}$, implying that
$V_{200} \, \gsim \, [10 \, G \, H_0 \, (\Omega_0/\Omega_b)\, M_{\rm
disk}]^{1/3}$.  For the disk mass adopted above and the cosmological parameters
appropriate to s$\Lambda$CDM discussed in \S1, we find that
$$ V_{200} \, \gsim \, 130 {\rm \, km \, s}^{-1}. \eqno(2) $$
We emphasize that this is a {\it strict} lower limit to the mass of the halo
that surrounds the Milky Way, since experiments show that typically not more
than $80\%$ of the baryons within $r_{200}$ are actually accreted into the
central disk (Navarro \& White 1994, Navarro \& Steinmetz 1997). A similar,
albeit more stringent, constraint may be derived by comparing the angular
momentum of the Milky Way with those of s$\Lambda$CDM halos. Halos with $V_{200}
\le 150$ km s$^{-1}$ typically have specific angular momenta lower than that of
the Milky Way disk (Mo, Mao \& White 1998, Syer et al 1999)  and are therefore
unlikely hosts of the Galaxy. Indeed, baryons typically lose a significant
fraction of their angular momentum as they collapse to the disk (Navarro \&
White 1994, Navarro \& Steinmetz 1997) so we conclude that almost certainly the
circular velocity of the Milky Way halo must exceed $130$-$150$ km s$^{-1}$.

Figure 1 compares the constraints from eqs.~1 and 2 with the results of our
s$\Lambda$CDM numerical simulations. The comparison shows clearly a major
discrepancy between the maximum dark matter inside $R_{o}$ allowed by
observations and the results of the numerical experiments. For example,
s$\Lambda$CDM halos with circular velocities similar to that of the Milky Way
disk ($V_{200} \approx V_c(R_{o})=220$ km s$^{-1}$) have about {\it three
times} more dark mass inside the solar circle than inferred from
observations. Even for the extreme case where the halo has the strict minimum
circular velocity allowed by eq.~2, the simulations indicate an excess of more
than $50\%$ in the dark mass within $R_{o}$.

This serious discrepancy only worsens if we take into account that some extra
dark material may be drawn inside $R_{o}$ by the formation of the disk. A
rough estimate of the magnitude of this correction can be made by assuming that
the halo responds adiabatically to the assembly of the disk; the discrepancy
then increases from $50\%$ to almost $80\%$ for the least massive halo allowed
by eq.~2. We conclude that halos formed in the s$\Lambda$CDM scenario are too
centrally concentrated to be consistent with observations of the dynamics of the
Galaxy.

\subsection{The Tully-Fisher relation}

It is possible to extend the analysis of the previous subsection to a large
fraction of disk galaxies by examining the tight correlation between the total
luminosity of galaxy disks and the rotation speed of their gas and stars (the
Tully-Fisher relation, Tully \& Fisher 1977). Provided that disk mass-to-light
ratios and exponential scalelengths can be estimated reliably, it is possible to
evaluate the disk contribution to the circular velocity and to apply the same
analysis of the previous subsection to derive constraints on the total dark mass
contained within the optical radius of the galaxy.

We choose to carry out the analysis at $2.2$ exponential scalelengths ($2.2 \,
r_{\rm disk}$) from the center, since the contribution of exponential disks to
the circular velocity peaks there and it is at that radius that Tully-Fisher
velocities are typically measured (Courteau 1997). (We note that $R_{o}\approx
2.4 r_{\rm disk}$ in the case of the Milky Way, so this choice of radius is
similar to that adopted in \S2.1.)  Constraints on dark masses inside $2.2 \,
r_{\rm disk}$ depend sensitively on estimates of the total mass associated with
galaxy disks. Because of the lack of an Oort-limit analog in external galaxies,
we resort to broad-band colors as estimators of the mass-to-light ratio of the
stellar disk. Late type spirals such as those that make up the majority of
galaxies in Tully-Fisher samples have $B-R$ colors in the range $(0.2,1.0)$
(Courteau 1999) which, for a galaxy that has been steadily forming stars for
$\sim 13$\, Gyrs, imply $I$-band mass-to-light ratios of order $(M_{\rm
disk}/L_I) \approx 2 \pm 1 (M/L_I)_{\odot}$. This estimate assumes a Salpeter
initial mass function and exponential star formation histories with timescales
which vary from $\tau_{\rm SF} \sim 1$ Gyr to star formation rates that are
constant over the age of the universe.  Similar values are obtained using the
GISSEL96 models of Bruzual \& Charlot (1996) and the PEGASE models of Fioc \&
Rocca-Volmerange (1997).  Uncertainties in this estimate of $(M_{\rm disk}/L_I)$
are hard to assess, but are unlikely to be larger than about a factor of two,
which is the value of the ``error'' we assume here. \footnote{We note that the
same procedure gives $V$-band mass-to-light ratios of about $2 (M/L_V)_{\odot}$
for galaxies in the same color range, more than a factor of $2$ lower than that
derived for the Milky Way disk from the Oort limit (Binney \& Tremaine
1987). This reflects the well known Galactic ``disk dark matter''
content. Assuming that Tully-Fisher galaxies have similar amounts of (presumably
baryonic) ``disk dark matter'' would make all the arguments we develop here even
stronger.}

With this caveat, we compute the rotation speeds ($V_{\rm rot}=V_c(2.2 \,
r_{disk})$) of hypothetical exponential disk galaxies with $M_{disk}/L_I=2 \,
(M/L_I)_{\odot}$ assembled at the center of the simulated dark halos shown in
Figure 1. Disk radii are chosen so as to satisfy the empirical relation, $r_{\rm
disk} \approx 3 \, (V_{\rm rot}/200$ km s$^{-1}) \, h^{-1}$ kpc (Courteau 1999,
Mo, Mao \& White, 1998, Navarro 1999).  Since disk radii depend on $V_{\rm
rot}$, an iterative procedure is needed, which we implement as follows. Given a
halo of circular velocity $V_{200}$ we assign to it a disk with exponential
scalelength derived assuming that $V_{\rm rot}=V_{200}$. The simulated halo
structure and the disk potential are then used to compute a new $V_{\rm rot}$
estimate, taking into account the ``adiabatic'' response of the halo to the disk
potential (see Navarro, Frenk \& White 1996, hereafter NFW96, for details). This
new velocity estimate is used to recompute $r_{\rm disk}$ and the procedure is
then iterated until convergence.

The solid-line curves in Figure 2 illustrate the result of applying this
procedure to three representative dark halos of different mass (or luminosity,
since we assume a constant mass-to-light ratio), as a function of the total mass
adopted for the disk of the galaxy. In each case we vary the total disk mass
from zero to $M_{\rm disk}^{\rm max}$, the maximum value compatible with the
baryonic content of the halo. As the disk mass increases, each hypothetical
galaxy moves from left to right across the plot. When the disk mass becomes
comparable to the dark mass inside $2.2 \, r_{\rm disk}$ the curve inches
upwards and becomes essentially parallel to the observed Tully-Fisher
relation. The increased potential due to the disk mass and the extra dark mass
drawn inside $2.2 \, r_{\rm disk}$ under the assumption that the halo responds
adiabatically to the assembly of the disk contribute similarly to the overall
increase in rotation speed over and above that of the original dark halo. The
rightmost point of each curve is reached when the mass of the disk equals
$M_{\rm disk}^{\rm max}$.

It is clear from Figure 2 that, even under the extreme assumption that galaxies
contain {\it all} available baryons in each halo, simulated disks are almost two
magnitudes fainter than observed. Increasing the baryonic mass of a halo has
virtually no effect on this conclusion, since in this case model galaxies would
just move further along paths approximately parallel to the Tully-Fisher
relation, as shown in Figure 2. Disk galaxies assembled inside s$\Lambda$CDM
halos therefore cannot match the observed Tully-Fisher relation, unless one or
more of the assumptions in our procedure are grossly in error.

Perhaps the most uncertain step in our argument is the stellar mass-to-light
ratio adopted for the analysis. The horizontal ``error bar'' shown on the
starred symbols in Figure 2 indicates the effect on our results of varying the
$I$-band mass-to-light ratio by a factor of two from our fiducial value of $2$
in solar units. This is not enough to restore agreement with observations, which
would require $(M_{\rm disk}/L_I) \sim 0.4$, a value much too low to be
consistent with standard population synthesis models. The vertical ``error
bars'' illustrate the effect of varying the ``concentration'' of each halo by a
factor of two.\footnote{The NFW ``concentration parameter'' is defined as
$c=r_{200}/r_s$, where the scale radius, $r_s$ is one of the parameters of the
density profile model proposed by NFW96 and NFW97, $\rho(r) \propto (r/r_s)^{-1}
(1+r/r_s)^{-2}$, see those papers for details.} Even with this large variation
in halo structure, our hypothetical disks fail to reproduce the observations.

A second uncertainty comes from the ``adiabatic contraction'' correction applied
to dark halos in order to mimic the halo response to the disk assembly, and one
may wonder whether our ``adiabatic'' contraction assumption is at all correct.
This possibility may be checked, however, through direct numerical simulation.
We have therefore included gas in the simulations and have evolved them again
using with the simplified treatment of radiative cooling and star formation
described in detail in Steinmetz \& Navarro (1999). The results are shown as
open squares in Figure 2, and are in good agreement with the results of the
simple modeling proposed above.

As in the case of the Milky Way discussed in \S2.1, the problem can be traced to
the large central concentration of s$\Lambda$CDM halos. Indeed, one way to solve
the problem would be to reduce substantially (by a factor of two or three) the
dark mass in the innermost few kpc of galaxy-sized dark halos. This would
significantly reduce $V_{\rm rot}$, bridging the gap between model and
observations. In terms of the halo density profile model proposed by NFW96 and
NFW97, this would be equivalent to reducing the ``concentration parameter'',
$c$, by a factor of about five.  This is the same conclusion reached by Navarro
(1999), who advocated that halos with very low ``concentration parameters'', $c
\, \lsim \, 3$, and stellar mass-to-light ratios as low as $\sim 0.5 \, h \,
(M/L_I)_{\odot}$ in low surface brightness galaxies were required in order to
match the shapes of disk galaxy rotation curves. We emphasize, however, that the
present conclusion is independent of assumptions about the detailed shape of the
dark matter density profile, and depends largely on the total dark mass on
scales of individual galaxies.

\section{Discussion and Conclusions}

The analysis of the previous subsection demonstrates conclusively that the
central mass concentration of s$\Lambda$CDM halos is inconsistent with
observations of the dynamics of spiral galaxies. Does this argument rule out all
models based on the CDM+inflation paradigm or are there plausible modifications
to the s$\Lambda$CDM parameters that may bring the model into agreement with
observations? 

In principle, there are parameter choices that may help make $\Lambda$CDM models
consistent with observations of the internal dynamics of galaxies, but the
magnitude of the modifications required are, however, uncomfortably large, and
come at the expense of other major successes of the model. As discussed by
NFW97, the central concentration of dark halos is directly proportional to the
mean matter density of the universe at a suitably defined collapse time.  For
halos of fixed mass, these collapse times may depend sensitively on the values
adopted for the cosmological parameters. However, as discussed by Navarro (1999,
see his Figure 8), the combination of parameters needed to reproduce the
present-day abundance of galaxy clusters is such that the characteristic
densities of galaxy-sized dark halos is approximately independent of $\Omega_0$
and of the value of $H_0$. Indeed, halos in the former ``standard'' CDM model
($\Omega_0=1$, $h=0.5$, $\sigma_8\sim 0.6$) have very similar concentrations as
the s$\Lambda$CDM halos we discuss here since both models are normalized to
match the abundance of massive clusters at $z=0$. This implies that all CDM
models that match the abundance of clusters are likely to have difficulty
reproducing the dynamics of spiral galaxies.

Even if cluster normalization is dropped from the list of relevant constraints
and only the COBE measurements are used to normalize the power spectrum of
initial density fluctuations, reducing substantially the central concentrations
of dark halos still implies uncomfortable parameter choices. For example, if
$\Omega_0$ is taken as a free parameter, reducing the dark mass of a $V_{200}
\sim 200$ km s$^{-1}$ halo inside $R_{o} \sim 8.5$ kpc by a factor of three
relative to s$\Lambda$CDM requires $\Omega_0 \sim 0.05$, a very low number that
would call into question the need for a dominant non-baryonic dark matter
component in the Universe. Similarly, if rather than $\Omega_0$, $H_0$ is
allowed to vary, values as low as $30$ km s$^{-1}$ Mpc$^{-1}$ are required to
obtain the desired effect, in gross disagreement with current observational
estimates.

If we are to preserve the successes of the s$\Lambda$CDM model, what seems to be
required is a substantial change in the {\it shape} of the power spectrum
relative to that predicted by the CDM+inflation paradigm. A ``designer'' power
spectrum that would go some way towards reconciling the model with observations
would suppress power on galactic and subgalactic scales while keeping the large
scale properties of the model virtually unchanged, as envisioned, for example,
in models that introduce a ``tilt'' in the primordial power spectrum relative to
the standard Harrison-Zel'dovich value.  This would in principle allow
galaxy-sized dark halos to collapse later and thus become less centrally
concentrated, although the magnitude of the tilt required is still
unclear. Other alternative modifications that may in principle reproduce the
desired trend would involve the existence of a sizeable ``hot dark matter''
component (although much higher than derived from current estimates of the
neutrino mass, see, e.g., Primack \& Gross 1998), or possibly even dark matter
candidates that may annihilate without trace in dense regions such as the
centers of galaxies. One problem that afflicts all these proposed modifications
is that they may hinder the formation of massive galaxies at high redshift, at
odds with the mounting evidence that such galaxies are fairly common at $z \,
\gsim \, 3$ (see, e.g., Steidel et al 1998). A thorough investigation of these
possibilities will be needed in order to assess just how radically the new
``standard'' model of structure formation must be revised in order to bring
spiral galaxies into the realm of observations that are consistent with the
current paradigm of structure formation.

\acknowledgments This work has been supported by the National
Aeronautics and Space Administration under NASA grant NAG 5-7151 and by NSERC
Research Grant 203263-98. MS acknowledges useful discussions with Walter Dehnen
and Rob Olling. JFN acknowledges enlightening discussions with Simon White,
Carlos Frenk, Mike Balogh, and Mike Hudson, as well as the hospitality of the
Max-Planck Institut f\"ur Astrophysik, where this manuscript was written.


\clearpage
\begin{figure}
\plotone{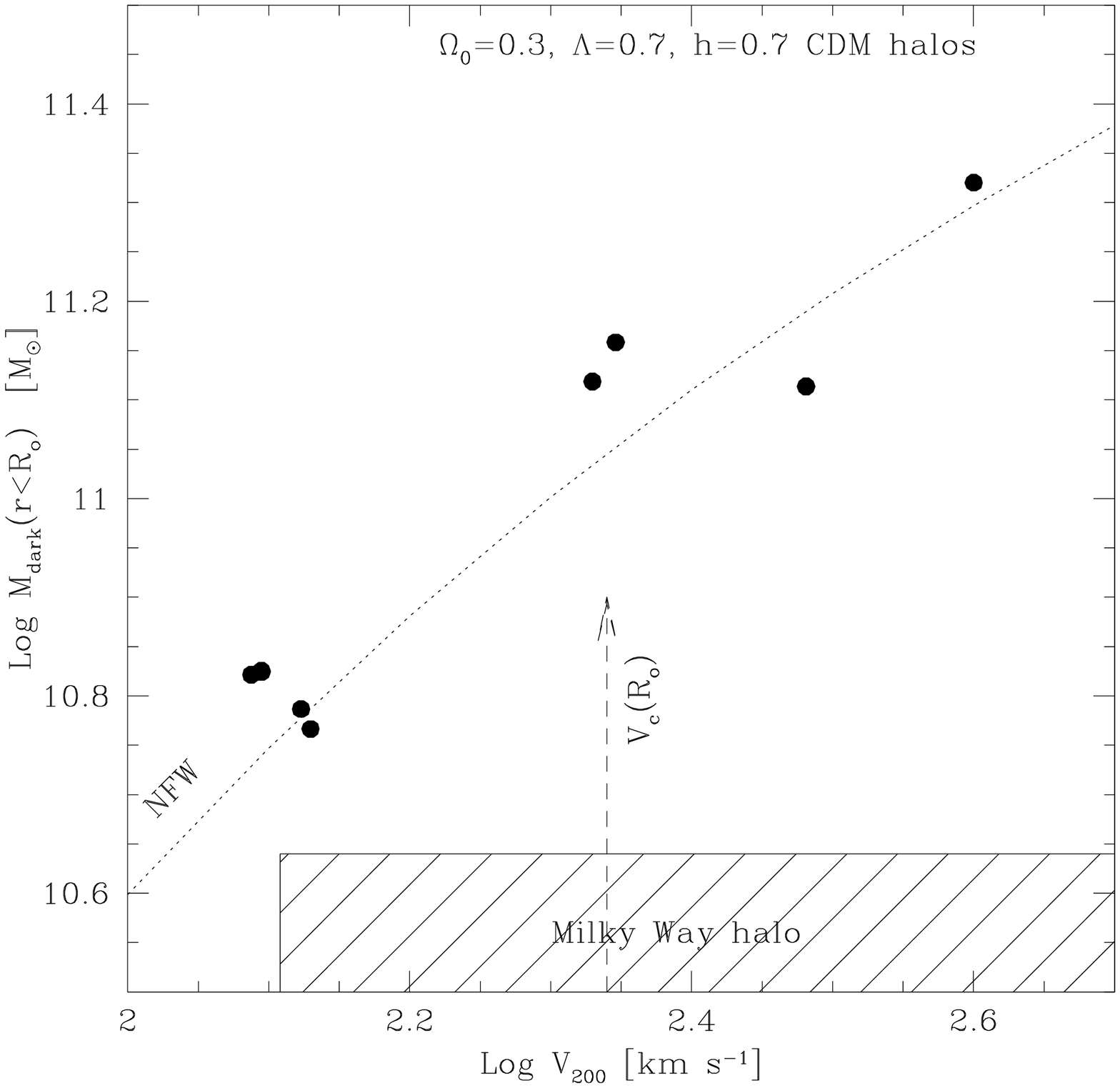}
\figurenum{1}
\caption{Dark mass enclosed within a radius $R_{o}= 8.5$ kpc, the
Sun's distance from the center of the Milky Way, versus the circular velocities
of s$\Lambda$CDM halos. The shaded region highlights the allowed parameters of
the dark halo surrounding the Milky Way, as derived from observations of
Galactic dynamics and by assuming that the disk mass cannot exceed the total
baryonic content of the halo. The filled circles show the loci of s$\Lambda$CDM
halos as determined from high-resolution N-body simulations. The solid line is
the circular velocity dependence of the dark mass expected inside $R_{o}$
for halos that follow the density profile proposed by NFW96 and NFW97. The
circular velocity dependence of the NFW ``concentration'' parameter of the
simulated halos is well approximated on these scales by $c \approx 20 \,
(V_{200}/100 {\rm \, km \, s}^{-1})^{-1/3}$ (dotted line). This is slightly higher
than predicted by the approximate formula proposed by NFW97 but consistent with
their published results.}
\end{figure}
\clearpage
\begin{figure}
\plotone{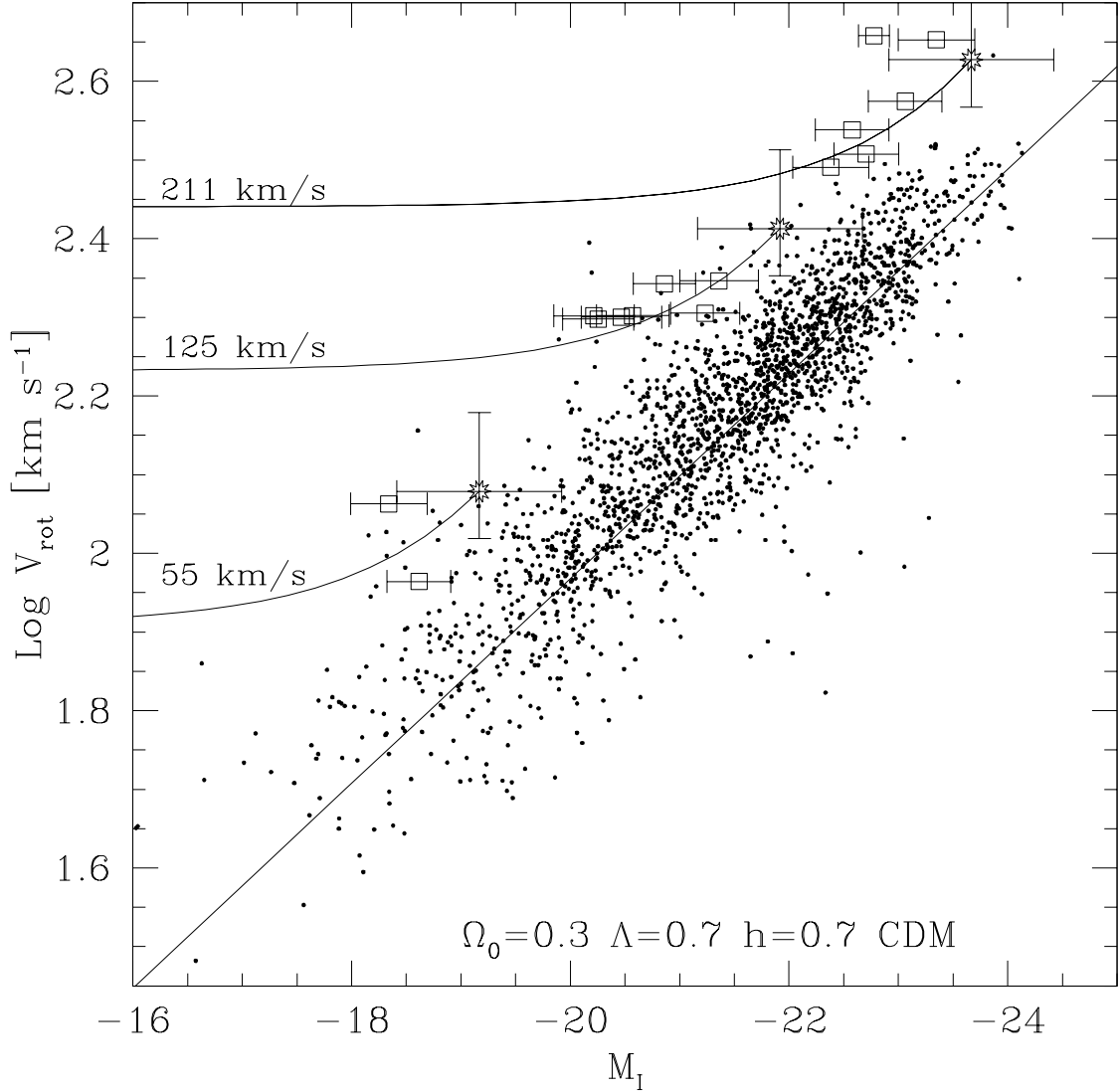}
\figurenum{2}
\caption{The $I$-band Tully Fisher relation compared with the loci of
hypothetical exponential disk galaxies assumed to assemble at the center of
three representative s$\Lambda$CDM halos. Dots are a compilation of the data by
Giovanelli et al.~(1997), Mathewson, Ford \& Buchhorn (1992) and Han \& Mould
(1992). The solid line is the best fit to the data advocated by Giovanelli et
al. The hypothetical galaxies have radii consistent with observations and move
from left to right along each curve (labeled by the circular velocity of the
halo at the virial radius) as the disk mass increases, under the assumption of a
constant stellar mass-to-light ratio, $M/L_I=2$ in solar units. The starred
symbols correspond to the maximum disk mass, $M_{\rm disk}^{\rm max}$, allowed
by the universal baryon fraction of the s$\Lambda$CDM model. Open squares are
N-body gasdynamical simulations of the formation of galaxies within these
halos. Error bars correspond to two different choices of IMF, as discussed in
detail by Steinmetz \& Navarro (1998). See that paper for details.}
\end{figure}
\end{document}